\documentclass{aipproc}
\layoutstyle{6x9}
\usepackage{graphics}
\def\lamAr{{\hat\lambda {\bf A} {\bf r}}}

\begin{document}

\title{Identifying Biomagnetic Sources in the Brain by the 
Maximum Entropy Approach}

\classification{87.10+e; 87.57.Gg}

\keywords{Magnetoencephalography, ill-posed problem, maximum entropy}

\author{Hung-I Pai}{
  address={
Department of Physics and $^\dagger$Department of Life Sciences\\
Computational Biology Laboratory\\
National Central University, Chungli, Taiwan 320, ROC}
}

\author{Chih-Yuan Tseng}{
  address={
Department of Physics and $^\dagger$Department of Life Sciences\\
Computational Biology Laboratory\\
National Central University, Chungli, Taiwan 320, ROC}
}

\author{HC Lee$^\dagger$}{
  address={
Department of Physics and $^\dagger$Department of Life Sciences\\
Computational Biology Laboratory\\
National Central University, Chungli, Taiwan 320, ROC}
}

\begin{abstract}
Magnetoencephalographic (MEG) measurements record magnetic fields
generated from neurons while information is being processed in the
brain. The inverse problem of identifying sources of biomagnetic fields  
and deducing their intensities 
from MEG measurements is ill-posed when the number of field detectors 
is far less than the number of sources.  This problem is less 
severe if there is already a reasonable prior knowledge in the form 
of a distribution in the intensity of source activation. 
In this case the problem of identifying and deducing source 
intensities may be transformed to one of 
using the MEG data to update a prior
distribution to a posterior distribution.  
Here we report on some work done using the maximum entropy method (ME)
as an updating tool. Specifically, we propose an implementation of the 
ME method in cases when the prior contain almost no knowledge of source  
activation. Two examples are studied,  in which part of
motor cortex is activated with uniform and varying intensities,
respectively.   

\end{abstract}

\maketitle

\section{Introduction}

Magnetoencephalographic (MEG) measurements record magnetic fields generated
from small currents in the 
neural system while information is being processed in
the brain \cite{Hamalainen}. In 
the classical cortical distributed model, the activation 
of neurons in the cortex is represented by sources of currents 
whose distribution approximates the cortex structure, and 
MEG measurements provide information on the current distribution  
for a specific brain function \cite{Hamalainen}. In practice, given an 
set of  current sources $J_{\beta}$, $\beta =\{1,\ldots ,N\}$ and a 
set of magnetic field detectors labeled by $l=\{ 1,\ldots ,d\}$, 
the relation between the field strengths $M_{l}$ measure by the 
detectors and the sources can be expressed as 
\begin{equation}
\label{ill-posed eq}
M_{l}=\sum\nolimits_{\beta=1}^{N} A_{l}^{\beta}J_{\beta}+\chi;\qquad 
l=1,\ldots, d,\ \beta =1,\ldots, N.   
\end{equation}
where {\bf A} is a $d\times N$ matrix whose elements $A_{l}^{\beta}$ 
are known functions 
of the geometric properties of the sources and the detectors, as 
determined by the Biot-Savart law \cite{Magnetic}, 
and $\chi$ indicates noise, to be ignored here.  The detail form 
of $A$ applicable to the present study is given in \cite{Bai03}.  
In tensor analysis notation 
Eq.~(\ref{ill-posed eq}) may be simply expressed as 
${\bf M} = {\bf A}\cdot{\bf J}$. 
In what follows, we adopt the convention of summing over 
repeated index ($\beta$ in Eq.~(\ref{ill-posed eq})). 
In standard MEG Eq.~(\ref{ill-posed eq})
appears as an inverse problem: the measured field strengths {\bf M} 
are given and the unknowns are {\bf J}. 
Since the total number $d$ of detectors that can be deployed in 
a practical MEG 
measurement is far less than the number of current sources, 
the answer to Eq.~(\ref{ill-posed eq}) is not unique 
and the inverse problem is ill-posed \cite{Clarke88,Alavi93}. 

A number of methods have been proposed to solve Eq.~(\ref{ill-posed eq}), 
including the least-square norm \cite{Hamalainen}, 
the Bayesian approach \cite{Baillet01,Schmidt97}, and the   
maximum entropy approach 
\cite{Clarke88,Alavi93,Khosla97,He00,Besnerais99,Amblard04}. 
In the method of maximum entropy (ME) the MEG data, in the form of 
the constraints {\bf M} - {\bf A}$\cdot{\bf J}=0$,  
is used to obtain a {\it posterior} 
probability distribution for 
neuron current intensities from a given {\it prior} (distribution). 
In \cite{Amblard04}, the method is implemented by introducing  
a hidden variable denoting the grouping 
property of firing neurons. 
Here we develop an approach such that ME becomes 
a tool for updating the probability 
distribution \cite{ShoreJohnson,JSkilling88,Caticha,Tseng04}.

\section{Computational detail}


\noindent {\bf ME updating procedure}. 
Let the set {\bf r} to be current intensities 
$r_\beta$ caused by neuron activities in the cortex at sites 
$\beta=1,\ldots, N$, and $p_\beta(r_\beta)$ be the 
probability current intensity distribution at site $\beta$.  
Assuming the N current sources to be uncorrelated, we define 
the joint probability distribution as
$P({\bf r}) = \prod_{\beta=1}^N p_\beta(r_\beta)$. The current 
at site $\beta$ is then $J_\beta =\langle r_\beta \rangle 
=\int r_\beta P({\bf r}) dr_\beta$. 
Suppose we have prior knowledge about neuron activities expressed in 
terms of the joint prior
$u({\bf r}) = \prod_{\beta=1}^N u_\beta(r_\beta)$.  The implication is that 
would produce currents {\bf J} that does not satisfy Eq.~(\ref{ill-posed eq}) 
(here without noise). 
Our goal is to update from this prior to a
posterior $P({\bf r}) d{\bf r}$ that does satisfy Eq.~(\ref{ill-posed eq}). 
The 
ME method states that given $u({\bf r})$ and the MEG data, the preferred
posterior $P({\bf r}) d{\bf r}$ is the one that maximizes relative 
entropy $S[P,u]$,
\begin{equation}
S[P,\mu] =-\int d{\bf r}\mbox{ }P\left( {\bf r}\right) \ln \left(P\left(
{\bf r}\right)/u \left( {\bf r}\right) \right) 
\label{rela-S}
\end{equation}
subject to constraints Eq.~(\ref{ill-posed eq}). Here $P({\bf r})$ 
is given by the variational method,
\begin{equation}
\label{P(r)}
P({\bf r}) d{\bf r}= Z^{-1} u({\bf r}) \exp(-\lamAr) d{\bf r}, \quad
Z=\int d{\bf r}\mbox{ }u({\bf r}) \exp(-\lamAr)
\stackrel{\mathrm{def}}{=}e^{-F},
\end{equation}
where $\hat\lambda$ is a row vector of length $d$ whose  $l^{\mbox{th}}$ 
element $\lambda^l$ is the Lagrangian multiplier that
enforces the $l^{\mbox{th}}$ constraints in Eq.~(\ref{ill-posed eq}),
$ \lamAr = \lambda^{l} A_{l}^{\beta} \langle r_{\beta}\rangle$ 
and the last equality in Eq.~(\ref{P(r)})defines the quantity $F$. 
Because {\bf A} is known, $P({\bf r})$ is determined by 
$u({\bf r})$ and the $\lambda$'s. 
The elements of $\lambda$ are the solutions 
$\lambda ^{l}=\bar{\lambda}^{l}$ in 
\begin{equation}
\label{e:solution_lambda}
A_{l}^{\beta} \left.\langle r_{\beta}\rangle 
\right\vert_{\lambda^l=\bar{\lambda}^l}
=\left. - \partial\ln Z/\partial \lambda^l
\right\vert_{\lambda^l=\bar{\lambda}^{l}}
=\left. \partial F/ \partial \lambda^l \right\vert_{\lambda
^{l}=\bar{\lambda}^{l}} = M_l,\quad l= 1,\ldots, d. 
\end{equation}
This is the primal-dual attainment equation derived in \cite{Besnerais99}. 
Because Eq.~(\ref{e:solution_lambda}) is a non-linear equation in the 
$\lambda$'s, the search for the $\bar\lambda$'s is non-trivial. 

A standard approach is by iteration, specifically by successive 
steps of updating $P({\bf r})$.  To demonstrate this we simplify notation 
and write $\nu^{\beta}=\lambda^{l} A_{l}^{\beta}$, or simply 
$\hat\nu=\hat\lambda {\bf A}$. Expectation value of currents can then be 
calculated through
$\langle r_{\beta}\rangle = - \partial \ln Z/ \partial \nu^\beta$.
The updating process may now start with $P_{[0]}=u({\bf r})$ and 
a set $\hat\lambda_{[0]}$ and proceed with   
\begin{equation}
P_{[i]}\left( r\right) d{\bf r}={Z^{-1}_{[i]}} P_{[i-1]}(r) 
e^{-\hat\nu {\bf r}}d{\bf r},
\quad Z_{[i]} = \int P_{[i-1]}(r) e^{-\hat\nu {\bf r}} d{\bf r},
\label{e:updating}
\end{equation}
where $[i= 1,2,\cdots]$ denotes the $i^{\mbox{th}}$ 
updating step and $\nu=\hat\lambda_{[i-1]} A$.  At each step $\hat\lambda_{[i-1]}$ is updated to 
$\hat\lambda_{[i]}$ according to Eq.~(\ref{e:solution_lambda}). 
Formally the updating process converges at the solution 
of Eq.~(\ref{e:solution_lambda}), which is a fix-point $\hat\lambda_*$
of Eq.~(\ref{e:updating}): $\hat\lambda_{[i]} 
= \bar{\hat\lambda} = \hat\lambda_*$.  
Then the current intensities will be fix-points $\langle r_\beta\rangle_*$
such that ${\bf M} = {\bf M}_* = {\bf A}\langle {\bf r}\rangle_*$.
In practice the fix-point may not be reached with infinite 
accuracy within finite time, and the updating may be terminated 
when the quantity 
\begin{equation}
\label{Bmse}
B_{mse} \stackrel{\mathrm{def}}{=} 
-10~\ln \left( {\left\Vert {\bf M}_* - {\bf M}\right\Vert}^{2}
/ \left\Vert {\bf M}\right\Vert ^{2} \right)  
\end{equation}
attains a predetermined value.
It is important to stress that unless the prior $u({\bf r})$ properly 
reflects sufficient knowledge about neuron activities, there is no guarantee 
that the fix-point $\langle {\bf r}\rangle_*$ is closely related to 
the actual current intensities. 


\begin{figure}[h]
\label{fig1}
\includegraphics[width=2.5in]{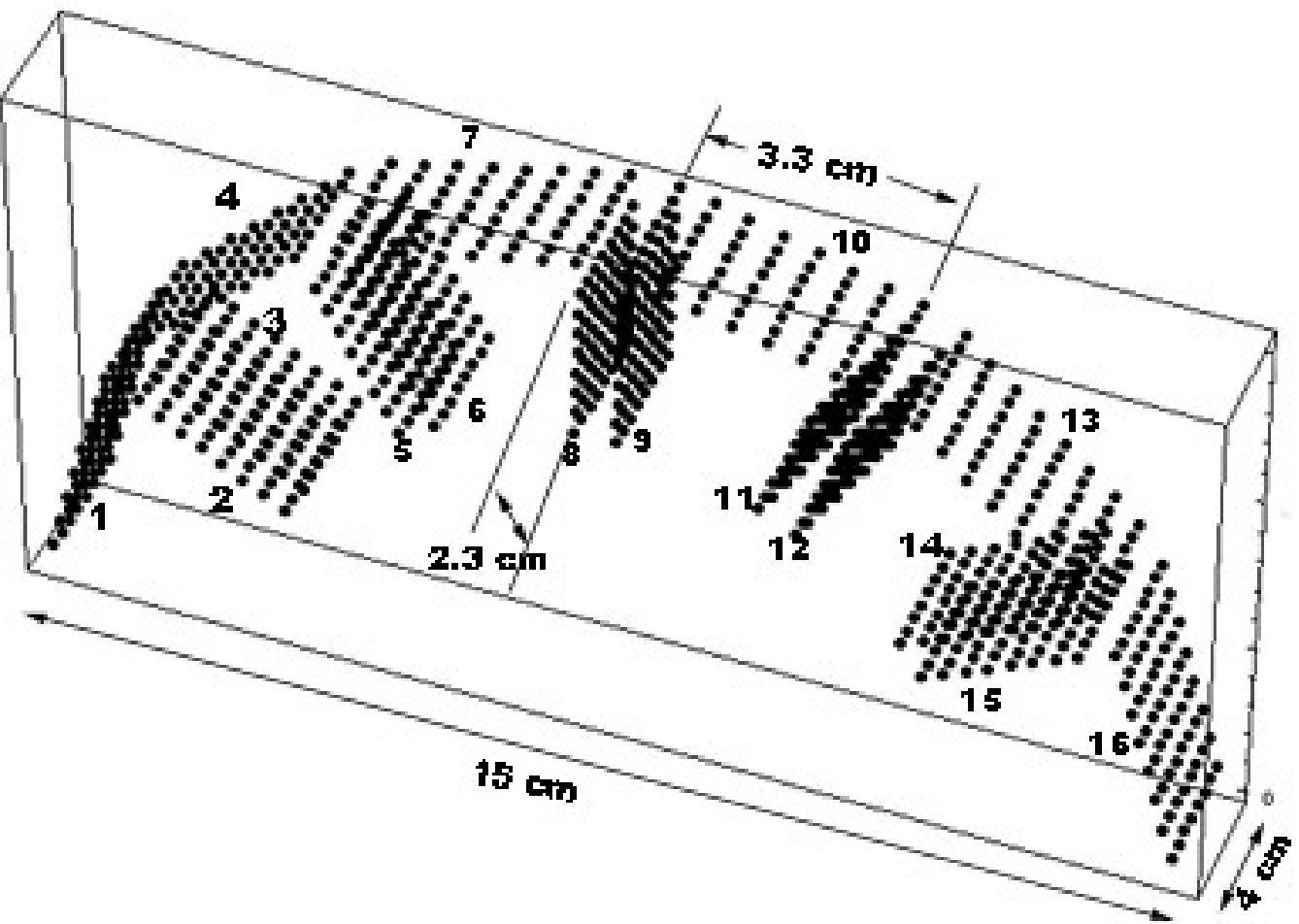}
\includegraphics[width=2.5in]{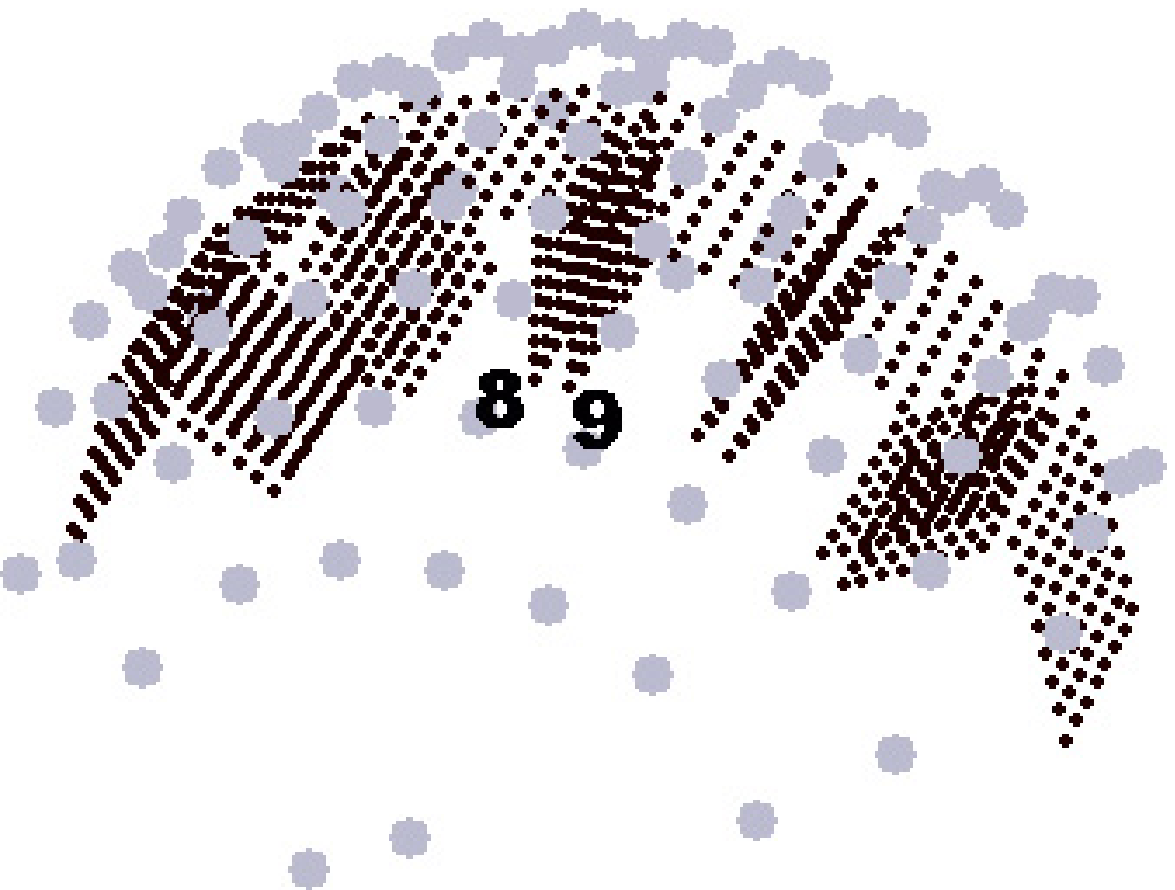}
\caption{A (left panel): Distribution of current sources used in study. 
The motor cortex is represented by the patches 7 to 10. 
B (right panel): Distribution of magnetic field detectors on a hemisphere 
2 cm from the scalp.}
\end{figure}
\smallskip\noindent {\bf Sources with Gaussian distributed intensities}. 
Pertinent general information on the geometric structure of the cortex 
and neuron activities, readily obtained from 
experiments such as functional
magnetic resonance imaging (fMRI), positron emission tomography (PET), 
etc., is incorporated in a 
distributed model \cite{Baillet01} in which current sources, 
modeled by magnetic dipoles, are distributed in regions below the scalp. 
A schematic coarse-grained 
representation of this model is shown in Fig.\ref{fig1}A,  
where 1024 dipoles are placed on 16 planar patches, 64 dipole to a patch. 
Eight of the which are parallel to the 
scalp and the other eight normal.  Regardless of the orientation of the 
patch, all dipoles are normal to the cortical surface with the 
positive direction pointing away from the cortex.  

Information contained in a 
prior may be qualitative instead of quantitative.  
Here, our prior will 
include the information that the activation resides in a part of the 
motor cortex that in Fig.\ref{fig1}A is represented by the patches 8 and 9, 
and utilize this prior information by placing a higher concentration 
of field detectors in the area nearest to those in Fig.\ref{fig1}B. 
 There is additional information such as neuronal grouping property. 
We follow Amblard {\it et al.} \cite{Amblard04} and  
group dipoles into cortical regions $C_{k}$, $k=1,2,\ldots K$, each 
containing $n_{k}$ dipoles with the $n_{k}$'s satisfying 
$N=\sum_{k=1}^{K}n_{k}$.
Associated with $C_{k}$ is a hidden 
variable $S_k$ that expresses regional activation status:  $S_{k}$=1 
denotes an ``excitatory state'', or a state of out-going current; 
$S_{k}$=-1 denotes an ``inhibitory state'' (in-going current); 
$S_{k}$=0 denotes a ``silent state'' (no current). 
With this grouping, the prior $u({\bf r})$ reduces to a sum of  
probability distributions $u({\bf r},S)$ over all possible configurations 
of $S=\{ S_{1},\cdots S_{K}\}$:   
\begin{equation}
\label{e:u-general}
u({\bf r}) d{\bf r} = \sum\nolimits_{S}u({\bf r},S) d{\bf r} 
=\sum\nolimits_{S}\mu ({\bf r}\vert S) \pi ( S) d{\bf r}
=\prod\nolimits_{k=1}^{K}\sum\nolimits_{S_{k}}
\mu( {\bf r}_{k}\vert S_{k}) \pi ( S_{k}) d{\bf r},
\end{equation}
where ${\bf r}_{k}=\{ r_\eta \vert \eta\in C_k\}$ specifies the current 
densities of the sources in $C_k$; $\mu({\bf r}_{k}\vert S_{k})
= \prod_{\eta\in C_k}\mu (r_{\eta}\vert S_{k})$ 
is the conditional joint probability of the dipoles in $C_k$ 
being in state $S_k$ and having current densities ${\bf r}_k$; 
$\pi(S_{k})$ is the probability 
of the region $C_k$ being in activation state $S_k$. 
For $\mu ({\bf r}_{k}\vert S_{k})$ we adopt 
a Gaussian distribution 
for activated states \cite{Clarke88,Alavi93,Amblard04}:
\begin{equation}
\mu( {\bf r}_{k}\vert S_{k}\neq 0 ) =
\prod_{\eta\in C_k} \frac{1}{\sqrt{4\pi \sigma}}
\exp \left[-\frac{1}{2\sigma }
( r_\eta-\rho_{\eta[0]})^2 \right].
\label{prior-conditional gauss}
\end{equation}
For simplicity, all current distributions 
have the same standard deviation $\sigma$.  
Current sources at different sites have different mean 
intensities $\rho_{i[0]}$ whose signs also indicate the state of 
activation: positive for excitation and negative for inhibition. 
For silent states we let $
\mu({\bf r}_{k}\vert S_{k}=0) =\delta({\bf r}_{k})=
\prod_{\eta\in C_k}\delta(r_\eta)$. 
We also write $\pi(S_{k}\ne 0) =1- \pi(S_k=0)=\alpha_{k}$, 
where $0\le\alpha_k\le 1$. We thus have, 
\begin{equation}
\label{prior-generalgauss}
u_{G}({\bf r})  
= \prod\nolimits_{k=1}^{K}
\sum\nolimits_{S_{k}}\mu({\bf r}_{k}\vert S_{k})\pi(S_{k})   
= \prod\nolimits_{k=1}^{K}\left[( 1-\alpha _{k}) \delta ({\bf r}_k) 
+\alpha _{k}\mu ( {\bf r}_{k}\vert S_{k}\neq 0) \right],   
\end{equation}
where the subscript denote Gaussian distribution. 
This form simplifies the computation significantly.  
At the $i^{\mbox{th}}$ iteration we have:
\begin{eqnarray}
Z_{[i]}=\mbox{\ }\prod\nolimits_{k=1}^{K}
\left(\prod\nolimits_{j=1}^{i-1}Z_{k[j]}\right)^{-1}
\left[1-\hat{\alpha}_{k[i-1]}+\hat{\alpha}_{k[i-1]}
\prod\nolimits_{j=1}^{i}\exp \left( F_{C_{k}\left[j\right] }\right)\right],
\label{e:guassian_iteration}
\end{eqnarray}
\begin{equation}
\langle r_{\eta }\rangle _{[i]} = \hat{\alpha} _{k[i]}\rho_{\eta [i]},\quad
\rho_{\eta [i]}=\rho_{\eta [i-1]}-\sigma \nu_{\eta [i-1]},\quad(\eta\in S_k),
\label{e:finetune_r}
\end{equation}
\begin{equation}
\hat{\alpha}_{k[i]}= \hat{\alpha}_{k[i-1]} \left( 
\hat{\alpha}_{k[i-1]}+\left( 1-\hat{\alpha}_{k[i-1]}\right) 
\prod_{j=1}^{i}\exp \left( -\bar{F}_{C_{k}[j]}\right) \right)^{-1}, 
\label{e:finetune_alpha}
\end{equation}
where $\hat{\alpha}_{k[0] }=\alpha _{k}$ and $\hat{\alpha}
_{k[1]}=\alpha _{k[1]}$; 
$F_{C_{k}[i]} = \sum_{\eta\in C_k}(2\sigma)^{-1}\left[
\left( \rho_{\eta [i]}\right) ^{2}-\left( \rho
_{\eta [i-1]}\right) ^{2}\right]$.  

In the absence of any other prior information we take $\alpha$ to be 
a random number (between zero and one),  $\vert \rho_{\beta [0]}\vert$ 
to have a random value up to 20 nA, 
the maximum current intensity that can be generated in the brain, 
and $\sigma$ to be the mean of 
$\vert \rho_{\beta [0]}\vert$.  However, the inverse problem being 
ill-posed, and since the prior contains no activation information, 
the above strategy produces poor results as expected. 



\smallskip\noindent {\bf Better priors by coarse graining}. 
In the absence of prior information on activation pattern, one way to 
acquire some ``prior'' information from the MEG data itself is by 
coarse graining the current source.  Coarse graining reduces the 
severity of the ill-posedness because the closer the number of current 
sources to the number of detectors, the less ill-posed the inverse 
problem.  Within the framework of the ME procedure described above, 
coarse graining can be simply achieved by setting $\rho_\eta$ for 
all $\eta$ in a given region $C_{k}$ to be the same.  Here we choose 
to take an intermediate step that disturbs the standard ME procedure 
even less, by replacing the second relation in  
Eq.~(\ref{e:finetune_r}) by  
\begin{equation}
 \rho_{\eta [i]}=
\langle r_{\eta }\rangle _{[i-1]}-\sigma \bar{\nu}_{\eta [i-1]}.
\label{e:coarse-grain}
\end{equation}
Note that in Eq.~(\ref{e:finetune_r}) 
$\langle r_{\eta}\rangle=\alpha_k\rho_\eta$  
depends on the probability $\alpha_k$ common to region $C_k$, whereas  
$\rho_\eta$ does not explicitly.  By replacing $\rho_{\eta [i-1]}$ 
by $\langle r_{\eta }\rangle _{[i-1]}$ on the right hand side of 
Eq.~(\ref{e:coarse-grain}), we force the updated $\rho_\eta$ in 
each iteration to be more similar (although not necessarily identical). 
In practice we only use this modified ME to get information on 
the activity pattern, rather than the intensity, of the sources.  Let 
$\langle \bf r\rangle_{\rm c}$ be the 
current intensity set obtained after 
a convergence criterion set by requiring $B_{mse}\ge B_{mse}^{\{c\}}$
(Eq.~(\ref{Bmse})). 
We now define a better prior set of Gaussian means $\hat \rho_{\rm c}$, 
where, in units of nA,
\begin{equation}
\label{e:firing_pattern}
\rho_{\beta\mbox{c}}=\left\{
\begin{array}{c}
{\rm sign}(\langle r_\beta\rangle_{\rm c})\ 20,\quad 
\vert\langle r_\beta\rangle_{\rm c}\vert >2,\\
0, \qquad\qquad\qquad \vert\langle r_\beta\rangle_{\rm c}\vert \le 2.
\end{array}\right.  
\end{equation}
These quantities, together with the obtained probabilities 
$\hat \alpha_{k{\rm c}}$ for the regions $C_k$, define a prior 
probability $P^{\{c\}}(\bf r)$, which may then be fed into the 
standard ME procedure for computing $\langle {\bf r}\rangle$. 

This procedure may be repeated by requiring  $B_{mse}$ to be not less than 
a succession of threshold values, 
$B_{mse}^{\{c1\}}< B_{mse}^{\{c2\}}< B_{mse}^{\{c3\}}< \cdots$, 
such that  a successive level of better priors 
$\hat \rho_{\rm c1}$, $\hat \rho_{\rm c2}$, $\hat \rho_{\rm c3}$, $\dots$, 
and $P^{\{c1\}}(\bf r)$, $P^{\{c2\}}(\bf r)$, $P^{\{c2\}}(\bf r)$, $\dots$, 
may be obtained. Eventually a point of diminishing return is reached. 
In this work we find the second level prior is qualitatively better than 
the first, and the third is not significantly better than the second. 


\section{Results}

In the following two examples, the 1024 current courses are partitioned 
into 16 patches, eight (4 cm wide and 3.3 cm long) parallel and 
eight (4 cm wide and 2.3 cm deep) normal to the scalp (Fig.\ref{fig1}A). 
On each patch lies a 8$\times$8 rectangular array of sources that are 
divided into 16 four-source groups;  that is, $K$=256.   
The interstitial distances on the horizontal (vertical) patches are 
0.57 and 0.47 cm (0.57 and 0.33 cm), respectively. The distance 
between the adjacent vertical patches are normally 0.55 cm, but the 
distance $d_{89}$ will be varied for testing, see below.  
The detectors are arranged in a hemisphere surrounding the 
scalp as indicated in  Fig.\ref{fig1}B. The matrix {\bf A} of 
Eq.~(\ref{ill-posed eq}) is given in \cite{Bai03}. 
The ME procedure is insensitive to $\sigma$ in the range 5$<\sigma<$100. 
In the coarse graining procedure we set 
$B_{mse}^{\{c1\}}$=100 and  $B_{mse}^{\{c2\}}$=150.  As noted previously, 
coarse graining a third time did not produce meaningful improvement 
on the prior. In the two examples, artificial MEG data are generated by 
having the sources on patch have uniform and varied current intensities, 
respectively. 

\begin{figure}[h]
  \includegraphics[height=1.8in]{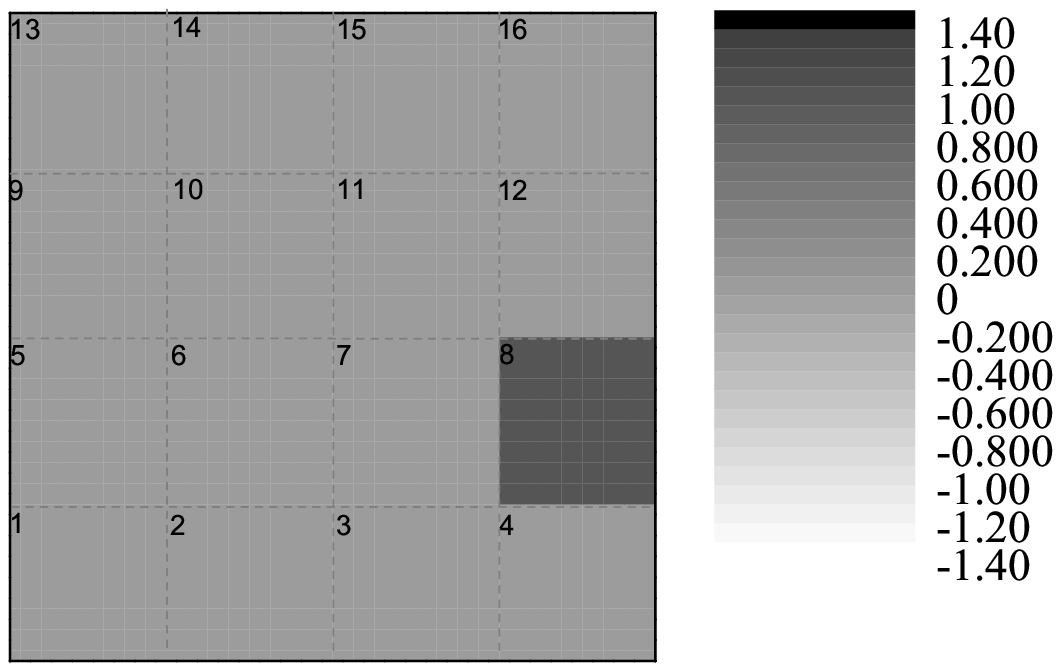}
 \includegraphics[height=2.0in]{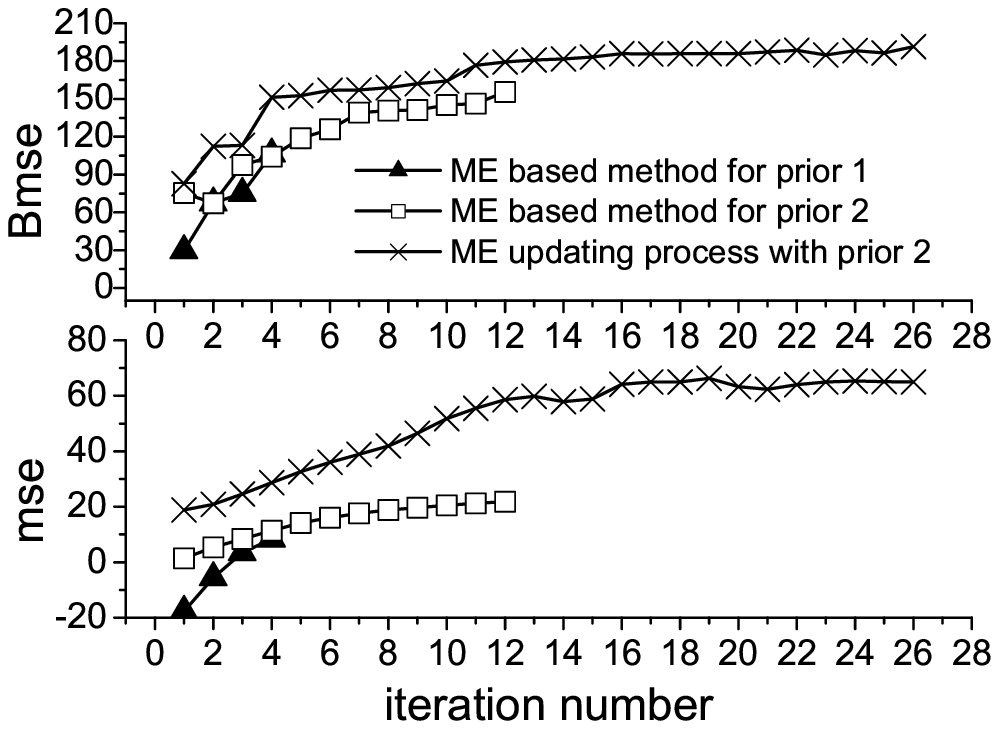}
 \caption{A (left panel): 
Contour plot indicates all 64 current sources on patch 8 
are activated with an intensity of 10 nA.  Gray scale on the right 
shows intensity level in units of 10 nA. 
Artificial MEG data {\bf M} used  in this section are generated 
through Eq.~(\ref{ill-posed eq}). 
B (right panel): $B_{mse}$ (top panel) and $mse$ (bottom panel) {\it vs}. 
iteration number.}
\label{fig2}
\end{figure}

\smallskip
\noindent {\bf Uniform activation on patch 8}. 
In this case the ``actual'' activity pattern is: the 64 sources on 
patch 8 each has a current of 10 nA, and all other sources are 
inactive (Fig.\ref{fig2}A.).  With the distance $d_{89}$ 
set to be 0.55 cm, the results 
in the first and second rounds of searching for a better prior, and 
in the final ME procedure proper are shown in Fig.\ref{fig2}B.  
In the plots,  $B_{mse}$ is the defined in Eq.~(\ref{Bmse}) and $mse$ 
is defined as
\begin{equation}
mse=-10\ln \left( \left\Vert \langle {\bf r}\rangle_{[i]}
- \langle{\bf r}\rangle \right\Vert ^{2}/
\left\Vert \langle{\bf r}\rangle \right\Vert ^{2}\right), 
\label{e:mse}
\end{equation}
where $\langle{\bf r}\rangle$ represents the actual source 
current intensity and the index $i$ indicates the iteration number.  
The solid triangles, squares, and 
crosses, respectively, give results from ME 
iteration procedures for constructing the first prior, 
second prior, and posterior.
It is seen that $B_{mse}$ rises rapidly in the search for the first prior 
(solid triangles); four iterations were needed for $B_{mse}$ to reach 100.  
$B_{mse}$ is less than 100 at the beginning of the second prior search 
because the prior values for this 
search is not the posterior of the previous search, but is related to 
it by Eq.~(\ref{e:firing_pattern}).  The same goes with 
the the relation between the beginning of the ME proper (crosses) and 
the end of second prior search (squares).  In the search 
for the second prior, $B_{mse}$ increases slowly after the seventh 
iteration, but eventually reaches 150 at the 12$^{\rm th}$ iteration.  
This already 
suggests that a round of search for a still better prior will not 
be profitable.  In the ME procedure proper, $B_{mse}$ reaches 150  
quickly at the fourth iteration, followed by a slow rise. After 
reaching 190 at the 14$^{\rm th}$ iteration the rise is very slow; the 
final value at the 26$^{\rm th}$ iteration is 195. 

The dependence of the $mse$ value on the ME procedures and the 
iteration numbers essentially mirrors that of $B_{mse}$.  The $mse$ 
for the final posterior is 68, which corresponds to an average of 
3.3\% error on the current intensities.

\begin{figure}[h]
\includegraphics[height=2.0in]{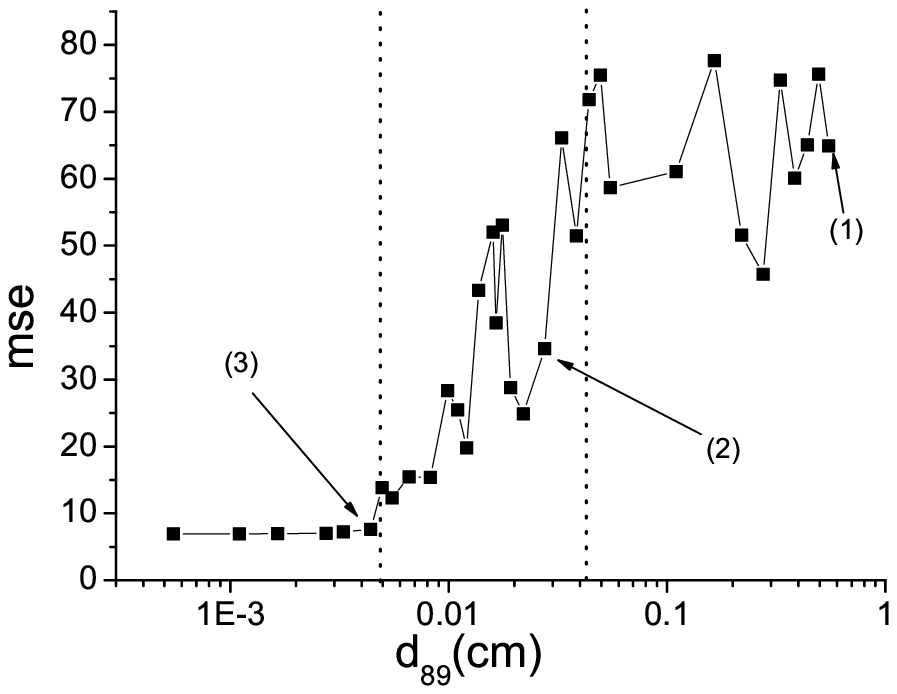}
\includegraphics[height=2.0in]{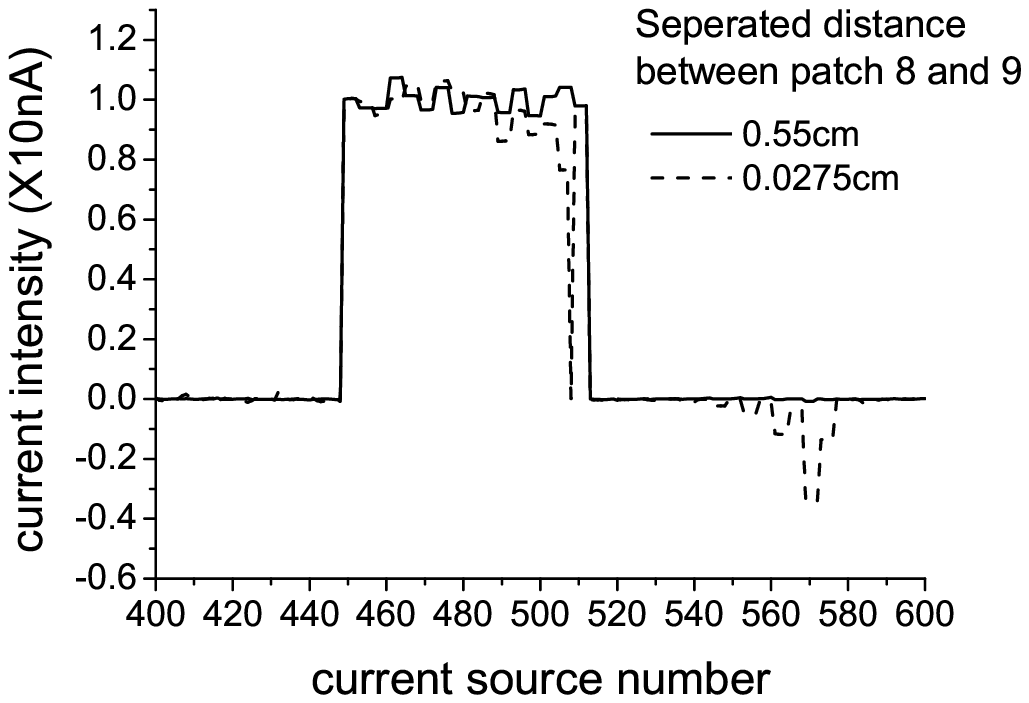}
\caption{A (left panel): $mse$ value {\it vs.} the separation 
$d_{89}$ between patches 8 and 9.  The distances $d_{89}$=0.55, 
0.0275 and 0.0035 cm are marked out and labeled (1), (2) and (3), 
respectively. 
B (right panel): 
Reconstructed $\langle{\bf r}\rangle$ {\it vs.} source number 
for the cases (1) (solid line) and case (2) (dashed line) in A.
}
\label{fig6_7}
\end{figure}

\noindent {\bf Resolving power as a function of $d_{89}$}.
We tested the resolving power of our ME procedure as a function of 
$d_{89}$. With 
uniform activation on patch 8, the computed $mse$ values
versus $d_{89}$ are plotted in Fig.\ref{fig6_7}A. 
The general trend is that $mse$ decreases with decreasing $d_{89}$ 
as expected: $mse = 60\pm 10$ when $d_{89}>$0.044 cm; 
$mse$ drops sharply when $d_{89}$ is less than 0.04 cm; 
$mse$ is less than 8 when $d_{89}$ is less than 0.0044 cm. 
In the last instance the ME procedure loses its resolving power because 
the error on the current intensity is about 70\%. 
On the other hand, $mse = 60\pm 10$ implies an error of 5.6$\pm$2.6\%. 
This means that if an error of no more than 8\% is acceptable, the 
ME method should be applicable to a source array whose density 
is up to one hundred times higher than that used in the present study. 


\smallskip\noindent {\bf Resolving power as a function of depth}. 
Signals from sources deeper in the cortex are in general weaker 
at the detectors and are harder to resolve. This effect 
is shown in Fig.\ref{fig6_7}B. The abscissa gives the source numbers 
on patch 8 (449 to 512) and patch 9 (513 to 576). The sources are 
arranged in equally spaced rows of eight, 
such that 449-456 and 513-520 are just below the scalp, 
457-464 and 521-528 are 0.328 cm from the scalp, and so on. 
Fig.\ref{fig6_7}B shows that when $d_{89}$=0.55 cm (solid line), the ME 
procedure can resolve all sources (up to a maximum depth of 2.3 cm). 
This resolving power decreases with decreasing $d_{89}$.  When 
$d_{89}$=0.0275 cm the ME procedure fails for sources at a depth of 
2 cm or greater (that is, sources 496-512 and 561-576 on patches 
8 and 9, respectively). 
\begin{figure}[h]
\includegraphics[height=2.1in]{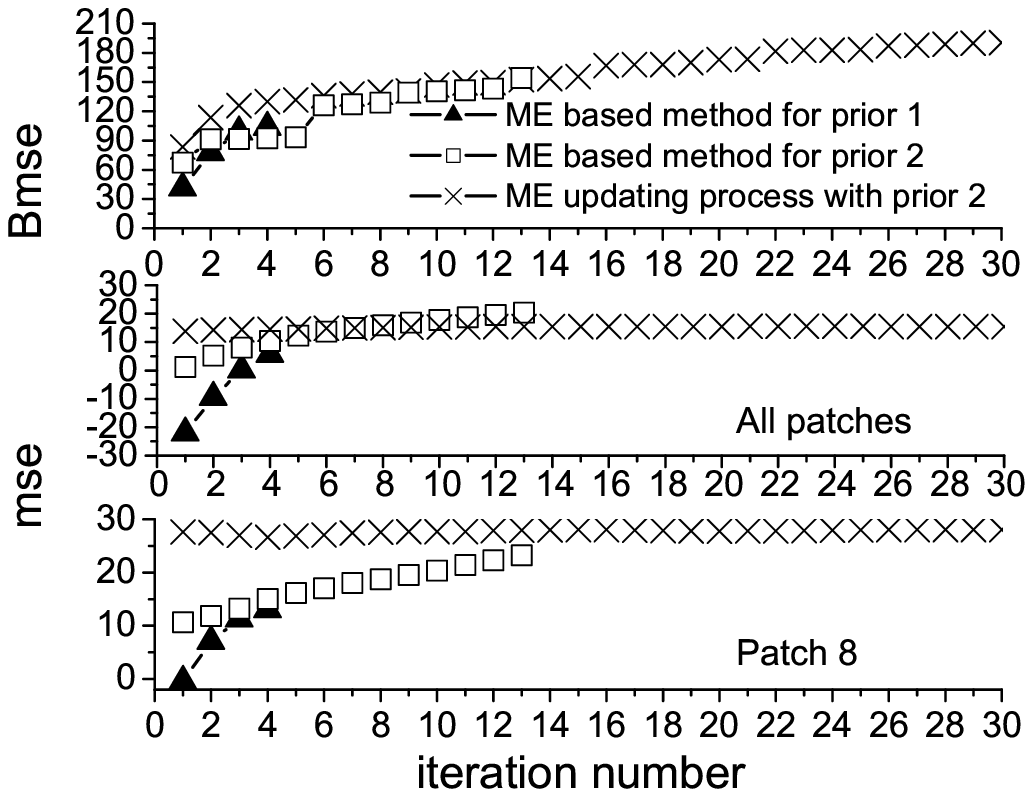}
\includegraphics[height=2.1in]{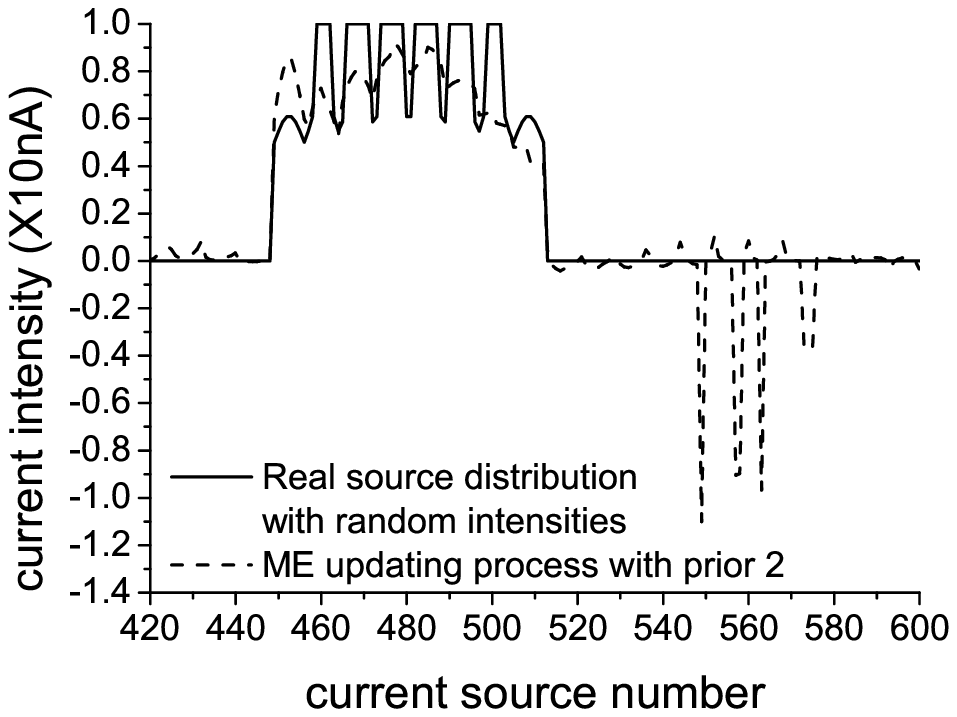}
\caption{A (left panel): $Bmse$ and $mse$ values for the case 
of varied activation on patch 8 (see text). 
B (right panel): Reconstructed (dash line) and (artificially generated) 
actual (black line) $\langle{\bf r}\rangle$  
$vs$. current source number on patches 8 and 9.}
\label{fig8}
\vspace{-1cm}
\end{figure}


\smallskip\noindent {\bf Varied activation on patch 8}.
We tested the ME procedure in a case with a slightly more 
complex activation pattern (still unknown to the prior): 
with $d_{89}$=0.55 cm, all current sources on patch 8 are activated, 
with those near the center of the patch having higher 
intensities than those in the peripheral.  All other sources are silent.   
We used random source current densities as zeroth order prior, employed 
coarse graining twice,  then used the standard ME to obtain the final 
reconstructed current intensities.  With $d_{89}$=0.55 cm,  
the dependence of $Bmse$ and $mse$ on the iteration number is shown in 
Fig.\ref{fig8}A.  Interestingly for the ME procedure proper 
(crosses), only the $Bmse$ improves with iteration, whereas the $mse$ 
value remains a constant at about 18.  This value is 
large compared to the value of 60$\pm$10 obtained for the case of 
uniform activation (Fig.\ref{fig2}B).  
The solid and dashed lines in Fig.\ref{fig8}B indicate the 
actual and reconstructed current intensities, respectively, for 
the sources on patches 8 and 9.  These show that the poor 
result is caused by reconstructed false activation of sources 550 to 580 
on patch 9.  
When sources other than those on patch 8 are forcibly forbidden to 
activate, $mse$ increases to 30 (bottom plot in Fig.\ref{fig8}A), 
corresponding to a 22\% error, suggesting   
that better results may be obtained when more reliable 
priors are given.  This remains to be investigated. 

\bigskip
\noindent {\bf Acknowledgments}. 
This work is partially supported by grants NSC 93-2112-M-008-031 
to HCL and NSC 94-2811-M-008-018 to CYT  
from National Science Council, ROC.  We thank Jean-Marc Lina for his 
contribution during the early stages of this work. 

\end{document}